\documentstyle[preprint,tighten,aps]{revtex}

\begin{document}
\draft
\title{ 
\begin{flushright}
\vspace{-1cm}
{\normalsize MC/TH 98/14}
\vspace{1cm}
\end{flushright}
The Wilson Renormalisation Group Applied to the Potential in 
NN Scattering\footnote{Talk given at The 11th Summer School and Symposium 
on Nuclear Physics (NuSS'98) ``Effective Theories of Matter'', 
held in Seoul National University, June 23-26, 1998.}}
\author{Michael C. Birse,  Judith A. McGovern and \underline{Keith 
G. Richardson}}
\address{Theoretical Physics Group, Department of Physics and Astronomy\\
University of Manchester, Manchester, M13 9PL, UK\\}
\maketitle
\begin{abstract}
Nonrelativistic two-body scattering by a short-ranged potential is studied
using the renormalisation group. Two fixed points are identified: a
trivial one and one describing systems with a bound state at zero energy. The
eigenvalues of the linearised renormalisation group are used to assign a
systematic power-counting to terms in the potential near each of these fixed
points. The expansion around the nontrivial fixed point defines a power counting scheme which is equivalent to the effective-range expansion.
\end{abstract}
\begin{section} {Introduction.}

The suggestion by Weinberg~\cite{wein2} that effective field theory 
(EFT) techniques could be usefully applied to nuclear physics has generated 
much discussion.  Since observed features in NN scattering such as
bound and nearly-bound states indicate that naive perturbation theory 
will fail, it was proposed that power counting rules should instead
be applied to the potential (or irreducible amplitude).  This potential
should then be iterated in a Schr\"odinger or Lippmann-Schwinger 
equation.

Immediately however there is a complication: the potential defined
in this way is highly singular.  Singular potentials are an unavoidable 
consequence of using local meson-nucleon and nucleon-nucleon couplings 
in the Lagrangian, and one must therefore specify a regularisation and 
renormalisation scheme in order to obtain finite physical quantities.  
Many such schemes have been proposed 
\cite{afg,ksw,md1,md2,lep,rbm,bvk,pkmr,ksw2,geg,uvk,ch}, 
but inconsistencies 
in the results obtained when the scattering length is large (as is the 
case in $S$-wave NN scattering) have led to the 
conclusion that Weinberg's power counting fails in such systems. 
(For details and references see \cite{drp,mcb} and Section \ref{regsch}).

A different power counting has recently been suggested by Kaplan, 
Savage and Wise (KSW)\cite{ksw2} in the framework of the 
power divergence subtraction scheme (PDS).  It has also been 
obtained by van Kolck \cite{uvk} in a 
scheme independent approach.  Although this power counting seems
to avoid the problems just mentioned, its basis and relation to the 
other schemes remain unclear.

To try and understand the failure of Weinberg's power counting and 
the apparent success of the new power counting, we have used the method of 
the Wilson (or ``exact'') renormalisation group \cite{wrg} applied to a 
potential which is allowed to have a cut-off dependence \cite{bmr}.  
We implement this by
imposing a momentum cut-off on the integrals encountered when 
iterating the full potential, and demanding that observables 
calculated in this way are cut-off independent.

The potential is rewritten in units of the cut-off $\Lambda$.  If $\Lambda$ 
is allowed to vary, the potential must change to ensure that the theory
still has the correct long range behaviour.  The variation of $V$ with
$\Lambda$ is given by a renormalisation group (RG) equation.
As the cut-off is lowered to zero, all finite ranged physics 
is integrated out of the theory, and 
physical masses and energy scales no longer appear explicitly.  
In the limit, the (dimensionless) rescaled potential must therefore become 
independent of $\Lambda$, the only remaining scale.  The many 
possible limiting values of the potential are fixed points of 
the renormalisation group equation.

We find two kinds of fixed point: the first is trivial, corresponding to 
an identically zero potential (no scattering), and the second to potentials 
producing a bound state exactly at threshold.  

By looking at perturbations around the single fixed point of the first type, 
we find that it is stable and corresponds to Weinberg's power counting.  
This organisation of the potential is systematic in a natural situation
where perturbation theory is applicable, but not in the case of interest
(low energy NN scattering).

The simplest fixed point of the second type is unstable, and 
the power counting we find here closely matches that which was 
found by KSW.  Perturbations around this fixed point
generate terms in the effective range expansion, which has
long been used to parameterise low energy NN data. 
\end{section}

\begin{section} {The Scattering Equation}
If one considers integrating all degrees of freedom out of the theory 
except nucleon fields, the resulting interaction Hamiltonian contains only 
four-nucleon contact terms.  This Hamiltonian can be arranged as a series
of terms containing derivatives of increasing order.  The resulting potential 
has the following simple expansion:
\begin{equation} \label{eq:pot1}
V(k',k,p)=C_{00}+C_{20}(k^2+k'^2)+C_{02}p^2\cdots,
\end{equation}
where $k$ and $k'$ are relative momenta and $p=\sqrt{ME}$ is the 
corresponding on-shell value at energy $E$.  

We shall work with the reactance matrix $K$ since it has a simpler
relation to the effective range expansion than the scattering matrix $T$.
The Lippmann-Schwinger (LS) equation for the off-shell $K$ matrix 
corresponding to the potential (\ref{eq:pot1}) is given by
\begin{equation} \label{eq:lse}
K(k',k,p)=V(k',k,p)+\frac{M}{2\pi^2}{\cal{P}}
\int_{0}^{\infty}q^2dq\,\frac{V(k',q,p)K(q,k,p)}
{{p^2}-{q^2}},\nonumber
\end{equation}
where ${\cal{P}}$ denotes a principal value integral.  The reactance matrix
consequently satisfies standing wave boundary conditions.   

The inverse of the on-shell $K$-matrix differs from that of the on-shell
$T$-matrix by a term $iM p/4\pi$, which ensures that $T$ is unitary if $K$ is
Hermitian.   Observables are obtained from $K$ by expanding the inverse
of its on-shell value in powers of energy.

\begin{equation} \label{eq:ere}
\frac{1}{K(p,p,p)}
=-\frac{M}{4\pi}\left(-\frac{1}{a}+\frac{1}{2}r_{e}p^2
+\cdots\right).
\end{equation}
This is just the familiar effective range expansion
where $a$ is the scattering length and $r_e$ is the effective range.

\end{section}

\begin{section} {Regularisation Schemes} \label{regsch}

A natural approach to regularising this theory is to impose a momentum
space cut-off at a scale $\Lambda<\Lambda_0$ where $\Lambda_0$ is a scale
corresponding to underlying physics which has been integrated out of the
effective theory \cite{md1,md2}. 
This can be done either by introducing a form factor 
in the potential reflecting the non-zero range of the interaction, 
or a cut-off on the 
momenta of intermediate virtual nucleon states.  In the case of a sharp 
momentum cut-off, the divergent integrals which arise in solving the 
scattering equation are of the form
\begin{equation} \label{eq:int1}
I_n=\frac{M}{2\pi^2}{\cal{P}}\int_{0}^{\Lambda}dq\,\frac{q^{2n+2}}
{{p^2}-{q^2}},\nonumber
\end{equation}
where extra factors of $q^{2}$ in the numerator of the integrand occur
in loop integrals with insertions of momentum dependent factors from the 
potential.

Expanding this integral in powers of energy, it is seen to have $n+1$
power law divergences,
\begin{equation} \label{eq:int2}
I_n=\frac{M}{2\pi^2}\left[
-\frac{\Lambda^{2n+1}}{2n+1}
-\frac{\Lambda^{2n-1}}{2n-1}p^2
+\ldots
-\Lambda p^{2n}
+p^{2n}\:I(p)
\right],
\end{equation}
\end{section}
where $I(p)$ is a function of $p$ which is finite as 
$\Lambda\rightarrow\infty$;
\begin{equation} \label{eq:ip}
I(p)=\frac{p}{2}\ln{\frac{\Lambda+p}{\Lambda-p}}.
\end{equation} 
A sharp cut-off is used here to avoid unnecessary complications.
If a different choice was made, the divergences in the $I_n$
would appear with different numerical coefficients, and the structure of 
$I(p)$ would change. Neither of these modifications would affect the 
conclusions below.

If the potential (\ref{eq:pot1}) is truncated at a given ``order'' in the
energy/momentum expansion, it has an $n$-term separable form, and we can
obtain an explicit expression for the $K$-matrix, and hence the observables
in the effective range expansion.  The undetermined coefficients in the
potential may then be fixed by demanding that they reproduce the observed 
effective range expansion up to the same order. 

In this way, by including more and more terms in the potential, it is 
possible to determine whether the calculation is systematic.  By systematic, 
we mean that the inclusion of extra terms in the potential should not result
in large changes to the coefficients which have already been fitted.  
Satisfying this demand ensures that the resulting Hamiltonian is 
meaningful outside the process in which it is determined.

Two distinct cases have been identified.  When the coefficients in the
effective range expansion are natural, the cut-off $\Lambda$ can be chosen
in such a way that the calculation is systematic.  So long as 
\begin{equation} 
\Lambda<\!<\frac{1}{a}\sim\frac{1}{r_e},
\end{equation}
the requirements given above are satisfied.  In this weak scattering
regime, perturbation theory is valid, and Weinberg's power 
counting rules apply.  

The scattering length $a$ is unnaturally large in $S$-wave NN scattering
however:
\begin{equation}
a\simeq24\:\hbox{fm}>\!>r_e\sim \frac{1}{m_\pi},
\end{equation}
and in this case choosing $\Lambda<\!<1/a$ would lead to a 
(systematic) theory with an extremely limited range of validity.  
One might hope that the following choice for the cut-off could be useful:
\begin{equation}
\frac{1}{a}<\!<\Lambda<\!<\frac{1}{r_e},
\end{equation}
but now corrections to the coefficients in the potential resulting from 
addition of extra terms will contain powers of both $1/\Lambda a$ and 
$\Lambda r_e$.  This choice does not lead to a systematic expansion in 
either $\Lambda$ or $1/\Lambda$.
 
An alternative is to use dimensional regularisation (DR) in 
which the loop integrals are continued to $D$ dimensions.  In the 
minimal subtraction scheme ($\overline{MS}$) \cite{ksw}, any logarithmic 
divergence (pole 
at $D=4$) is subtracted and power law divergences do not appear.  Since 
no logarithmic divergences appear in (\ref{eq:int2}), the loop integrals 
$I_n$ are set to zero and the $K$-matrix is simply given by the first Born
approximation,
\begin{equation}
K(k',k;p)=V(k',k,p).
\end{equation}
From the on-shell version of this identity we can obtain the 
effective range expansion which is found to converge only in the 
region $p<\sqrt{2/ar_e}$. As in the case of a cut-off $\Lambda<\!<1/a$ 
this scheme is always systematic, but useless when $a$ is large.

Kaplan, Savage and Wise \cite{ksw2} have proposed using a different scheme 
called power-divergence subtraction (PDS) to avoid this problem.  PDS 
subtracts linear divergences (corresponding to poles at $D=3$), so
that the $I_n$ do not vanish.  The resulting $K$-matrix now has a 
dependence on the subtraction scale $\mu$, and choosing
\begin{equation}
\frac{1}{a}<\!<\mu\sim p<\!<\frac{1}{r_e},
\end{equation} 
where $p$ is the momentum scale of interest, a systematic power
counting scheme emerges which appears to be useful.  Note that the subtracted
term, which is linear in $\Lambda$, is the only divergence which occurs 
when the potential is momentum independent ($C_{2i,0}=0$ in (\ref{eq:pot1})).
PDS must therefore be interpreted as a momentum independent scheme.  We
shall return to this point later.
Gegelia \cite{geg} 
has obtained a similar result by performing a momentum subtraction at 
the unphysical point $p=i\mu$.  The results obtained using these schemes 
agree with those of van Kolck (in a scheme independent approach) \cite{uvk} 
and more recently Cohen and Hansen \cite{ch}
(in coordinate space).
In the next section, we shall see that this new power counting can be
understood in terms of an expansion around a fixed point of the  
renormalisation group (RG) equation which we derive for the 
potential.  We shall also make clear that it is, at least to all
orders considered so far, equivalent to an effective range 
expansion (as suggested by van Kolck \cite{uvk}).

\begin{section} {The Renormalisation Group}

The starting point for the derivation of the RG equation
is the LS equation (\ref{eq:lse})
where all quantities are now allowed to depend on $\Lambda$.  $V$ is
the exact potential required to reproduce the observed $K$-matrix, and will 
therefore be $\Lambda$ dependent after renormalisation  
in a cut-off scheme.  The $\Lambda$ dependence of the free Green's function
$G_0$ regulates divergent loop integrals.  As above, we use a sharp
cut-off $\Lambda$ on loop momenta.  This choice will simplify the
discussion but, as before, our results apply equally well to
any reasonable choice of momentum space cut-off.

To proceed, it is helpful to 
demand that the entire off-shell $K$-matrix is independent of 
$\Lambda$.  Note that this is a stronger requirement than is necessary 
simply to ensure cut-off independence of the resulting observables.
After differentiating the LS-equation with respect to $\Lambda$ and setting 
$\partial K/\partial\Lambda=0$, the RG-equation for 
$V(k',k,p,\Lambda)$ can be obtained by operating on the resulting expression
from the right with $(1+G_0K)^{-1}$ to get
\begin{equation}\label{eq:rge}
{\partial V\over\partial\Lambda} 
={M\over2\pi^2}V(k',\Lambda,p,\Lambda){\Lambda^2\over\Lambda^2-p^2}
V(\Lambda,k,p,\Lambda).
\end{equation}
In the derivations to follow, we will impose the boundary conditions
that $V$ should have an expansion in powers of $p^2$, $k^2$ and $k'^2$.
This means that it can be obtained from the type of Hamiltonian proposed by 
Weinberg \cite{wein2} and written in the form (\ref{eq:pot1}).

To bring out the interesting features of this approach, it is useful 
to introduce the dimensionless momentum variables, $\hat{k}=k/\Lambda$ etc.,
and the scaled potential
\begin{equation}
\hat V(\hat k',\hat k,\hat p,\Lambda)={M\Lambda\over 2\pi^2}
V(\Lambda\hat k',\Lambda\hat k,\Lambda\hat p,\Lambda),
\end{equation}
where the overall factor of $M/2\pi^2$ comes from the LS-equation 
(\ref{eq:lse}).  

Finally, we rewrite the RG-equation in terms of these new variables,
\begin{equation}\label{eq:scrge}
\Lambda{\partial\hat V\over\partial\Lambda}=
\hat k'{\partial\hat V\over\partial\hat k'}
+\hat k{\partial\hat V\over\partial\hat k}
+\hat p{\partial\hat V\over\partial\hat p}
+\hat V+\hat V(\hat k',1,\hat p,\Lambda){1\over 1-\hat p^2}
\hat V(1,\hat k,\hat p,\Lambda).
\end{equation}

In the following, the idea of a fixed point will
be important.  As $\Lambda$ varies, the RG-equation (\ref{eq:scrge})
describes how the rescaled potential must flow to ensure that the
long range behaviour of the theory does not change.  As the cut-off
is taken to zero, more and more  physics is integrated out of the 
theory.  In the limit, the cut-off is the only remaining energy scale and 
the rescaled potential, being dimensionless,  must become independent of 
$\Lambda$.  The many possible limiting values of the potential are fixed 
points of the renormalisation group equation.
To find fixed points, we look for solutions of (\ref{eq:scrge}) which satisfy
\begin{equation}
\Lambda\frac{\partial\hat V}{\partial\Lambda}=0.
\end{equation}
Near a fixed point, all physical energies and masses are large compared to 
the cut-off and power counting becomes possible.  

A simple example which turns out to be important is the trivial
fixed point $\hat V(\hat k',\hat k,\hat p,\Lambda)=0$. 
It is easy to see that the $K$-matrix calculated from this
potential is also zero, corresponding to no scattering.  
It is necessary to know how perturbations around the fixed point potential 
scale if we are to use it as the basis of a power counting
scheme.  This can be done by linearising the RG-equation by looking 
for eigenfunctions which scale as an integer power of $\Lambda$,
\begin{equation}
\hat V(\hat k',\hat k,\hat p,\Lambda)=
C\Lambda^\nu \phi(\hat k',\hat k,\hat p).
\end{equation}
Substituting this back into (\ref{eq:scrge}), the linear eigenvalue equation
for $\phi$ is found to be
\begin{equation}\label{eq:linrge.tr}
\hat k'{\partial\phi\over\partial\hat k'}
+\hat k{\partial\phi\over\partial\hat k}
+\hat p{\partial\phi\over\partial\hat p}
+\phi=\nu\phi.
\end{equation}
The solutions of this equation which satisfy the boundary conditions 
specified above are easily found to be
\begin{equation}
\phi(\hat k',\hat k,\hat p)=\hat k^{\prime l}\hat k^m \hat p^n,
\end{equation}
with RG eigenvalues $\nu=l+m+n+1$, where $l$, $m$ and $n$ are non-negative 
even integers.  The momentum expansion of the rescaled potential around the 
trivial fixed point is therefore given by 
\begin{equation}
\hat V(\hat k',\hat k,\hat p,\Lambda)=\sum_{l,n,m}\widehat C_{lmn}\left(
{\Lambda\over\Lambda_0}\right)^\nu \hat k^{\prime l}\hat k^m \hat p^n.
\end{equation}
(For a hermitian potential one must take $C_{lmn}=C_{mln}$.)  The coefficient
$\widehat C_{lmn}$ has been made dimensionless by taking out a factor 
$1/\Lambda_0^\nu$ where, as before, $\Lambda_0$ is a scale corresponding
to underlying (integrated out) physics.

Since all of
the allowed eigenvalues are positive, this potential vanishes as 
$\Lambda/\Lambda_0\rightarrow 0$, and the trivial fixed point is stable.  
The power counting scheme associated with this fixed point can be 
made explicit by considering the unscaled potential 
\begin{equation}\label{eq:potexp.tr}
V(k',k,p,\Lambda)={2\pi^2\over M\Lambda_0}\sum_{l,n,m}\widehat C_{lmn}
{k^{\prime l}k^m p^n\over\Lambda_0^{l+m+n}}.
\end{equation}
So long as the $\widehat C_{lmn}$ are natural, the contributions
of energy and momentum dependent terms to the potential are suppressed 
by powers of $1/\Lambda_0$.  Assigning an order $d=\nu-1$ to these
perturbations leads to a scheme which is exactly equivalent to that
originally suggested by Weinberg \cite{wein2}.

In section \ref{regsch}, we saw that power counting 
in this way is useful in the case where scattering is weak at
low energies, and either dimensional regularisation with minimal 
subtraction or a cut-off may be used.   When the scattering length 
is unnaturally large however, we are forced to choose $\Lambda$ 
to be so small that the theory becomes essentially useless.  To find a 
useful expansion of the potential in this case, it is necessary to 
look for a non-trivial fixed point which describes strong scattering
at low energies.

The simplest fixed point of this kind can be found by looking for
a momentum independent potential, $\widehat V=\widehat V(\hat p)$, 
which satisfies the full RG-equation.  This becomes  
\begin{equation}\label{eq:fprge.ere}
\hat p{\partial\hat V_0\over\partial\hat p}
+\hat V_0(\hat p)+{\hat V_0(\hat p)^2\over 1-\hat p^2}=0.
\end{equation}
Solving this equation subject to the boundary condition that the potential be 
analytic in $\hat p^2$ as $\hat p^2\rightarrow 0$ we obtain
\begin{equation}
\widehat V_0(\hat p)=-\left[1-{\hat p\over 2}\ln{1+\hat p\over 1-\hat p}
\right]^{-1}.
\end{equation}
The corresponding unscaled potential is 
\begin{equation} \label{eq:potfp.ere}
V_0(p,\Lambda)=-{2\pi^2\over M}\left[\Lambda-{p\over 2}
\ln{\Lambda+p\over\Lambda-p}\right]^{-1}=-{2\pi^2\over M}\left[\Lambda-I(p)
\right]^{-1},
\end{equation}
where $I(p)$ was defined in Eq. (\ref{eq:ip}).
The RG-equation (\ref{eq:scrge}) was derived using a sharp cut-off, and
this form for the fixed point potential is a consequence of that 
choice.  Choosing another type of cut-off would change the second 
($p$ dependent) term in the square brackets, but the $1/\Lambda$ 
behaviour as $p\rightarrow 0$ is independent of the form of cut-off used.

Inserting this potential in the LS-equation, we find $1/K=0$.  This
fixed point therefore corresponds to a zero energy bound state (i.e.
the scattering length $a$ is infinite).  
Not surprisingly, this fixed point has properties which are quite different
from those we found in the trivial case.
As before, we look at perturbations around this fixed point which 
scale as an integer power of $\Lambda$
\begin{equation} 
\hat V(\hat k',\hat
k,\hat p,\Lambda)=\hat V_0(\hat p)+C\Lambda^\nu \phi(\hat k',\hat k,\hat p),
\end{equation}
where $\phi$ is a function which is well behaved for small momenta and 
energy and satisfies the linearised RG-equation:
\begin{equation}\label{eq:linrge.ere}
\hat k'{\partial\phi\over\partial\hat k'}
+\hat k{\partial\phi\over\partial\hat k}
+\hat p{\partial\phi\over\partial\hat p}+\phi\\
+{\hat V_0(\hat p)\over 1-\hat p^2}\left[\
\phi(\hat k',1,\hat p)+\phi(1,\hat k,\hat p)\right]
=\nu\phi.
\end{equation}

The general case $\phi=\phi(\hat k',\hat k,\hat p)$ can be tackled by
breaking it down into simpler types of perturbation.  Perturbations 
which depend only on energy $\phi=\phi(\hat p)$ turn out to be particularly 
important, so we will consider these first.  Equation (\ref{eq:linrge.ere}) 
can be integrated subject to the same boundary conditions as before, 
and the result is simply 
\begin{equation}\label{enper.ere}
\phi(\hat p)=\hat p^{\nu+1} \hat V_0(\hat p)^2.
\end{equation}

with eigenvalues $\nu=-1,1,3,5,...$.  
The  first of these eigenvalues is negative, and is associated
with an unstable perturbation since it gives rise to a term in the 
potential which scales like $\Lambda_0/\Lambda$ as $\Lambda\rightarrow0$.
This fixed point is therefore unstable; only potentials which initially
lie exactly on the critical surface $1/K=0$ approach it as the cut-off 
is taken to zero.  

We must also consider perturbations which are momentum dependent.  This time, 
we look for a function $\phi$ of the form
\begin{equation}
\phi(\hat
k',\hat k,\hat p)=\hat k^n\phi_1(\hat p)+\phi_2(\hat p)
\end{equation}
which satisfies (\ref{eq:linrge.ere}).  The solutions are given by
\begin{equation}\label{eq:mompert.ere}
\phi(\hat k',\hat k,\hat p)=\left[\hat k^n-\hat p^n+\left({1\over n+1}
\right.\right.+{\hat p^2\over n-1}+\cdots\\
+\left.\left.{\hat p^{n-2}\over 3}\right)\hat V_0(\hat p)
\right]\hat V_0(\hat p).
\end{equation}
If a Hermitian potential is required, a matching perturbation of this form 
should be added with $k$ replaced by $k'$.  As usual, the
eigenvalues are given by the condition that the potential be 
a well behaved function of energy and squared momenta; in this 
case they are $\nu=n=2,4,6\dots$.

In general, new solutions to the linearised RG equation (\ref{eq:linrge.ere})
can be obtained by multiplying existing ones by $p^m$ where $m$ is a positive
even integer.  Applying this to the functions in equation (\ref{eq:mompert.ere})
results in a new set of eigenfunctions with corresponding eigenvalues 
$\nu=n+m$. 	

An important point to note is that the momentum-dependent
eigenfunctions have different eigenvalues from the corresponding purely
energy-dependent ones and so they scale differently with $\Lambda$.
This is quite unlike the more familiar case of perturbations around the trivial
fixed point where, for example, the $\hat p^2$ and $\hat k^2$ terms in the 
potential are both of the same order, $\nu=3$. It means that, 
in the vicinity of the nontrivial fixed point, one cannot use the 
usual equations of motion to eliminate energy dependence from the 
potential in favour of momentum dependence.

One more kind of perturbation is possible: a product of two factors 
of the type given in equation (\ref{eq:mompert.ere}), where one is a 
function of $\hat k$ and the other is a function of $\hat k'$, is a
solution of the linearised RG-equation with eigenvalues $\nu=5,7,9\dots$.  
These solutions can also be multiplied by powers of $p^2$ to give 
eigenfunctions with higher eigenvalues.

Any perturbation which solves equation (\ref{eq:linrge.ere}) subject
to the given boundary conditions can expressed as a sum of those
 given above,
so we are in a position to expand a general potential around this
fixed point.  Near the fixed point we need only  
consider the perturbations with the lowest eigenvalues.  Including
the unstable perturbation, the first three give the following unscaled
potential 
\begin{eqnarray} \label{eq:pot2}
V(k',k,p,\Lambda)=V_0(p,\Lambda)
&+&{M\Lambda_0\over 2\pi^2}\sum_{\nu=-1}^3\widehat C_\nu\left({p\over\Lambda_0}
\right)^{\nu+1}V_0(p,\Lambda)^2\\ \nonumber
&+&\widehat C_2\left[k'^2+k^2-2p^2+{2\over 3}{M\Lambda^3\over 2\pi^2}V_0\right]
{V_0(p,\Lambda)\over\Lambda_0^2}.
\end{eqnarray}
Although this potential looks complicated, it has a two term separable
structure, and we can obtain an exact expression for the $K$-matrix
\cite{md2,rbm}.  Expanding the inverse of the on-shell $K$-matrix, 
we find the following effective range expansion
\begin{equation}\label{eq:enere.ere}
\frac{1}{K(p,p,p)}=-{M\Lambda_0\over 2\pi^2}
\sum_{\nu=-1}^3\widehat C_\nu\left({p\over\Lambda_0}\right)^{\nu+1}+\cdots,
\end{equation}
to first order in the $\widehat C_\nu$.  

To this order, the potential is determined uniquely by comparing equation
(\ref{eq:enere.ere}) with the effective range expansion (\ref{eq:ere}). 
\begin{equation} \label{eq:coefts}
\widehat C_{-1}=-{\pi\over 2\Lambda_0 a},\qquad 
\widehat C_1={\pi\Lambda_0 r_e\over 4}\qquad \ldots
\end{equation}
The identification of the terms in the potential and the effective 
range expansion is straightforward at this order because only energy dependent 
perturbations contribute to the scattering.  It is not clear 
that this equivalence persists to higher order in the $\widehat C_\nu$.  
To illustrate this, we have calculated corrections to the unscaled potential 
(\ref{eq:pot2}) up to order $\widehat C_{1}\widehat C_{2}$.  We find the 
following extra contributions 
\begin{equation}
\frac{M}{2\pi^2}\widehat C_{3}
p^4\frac{V_0(p,\Lambda)^2}{\Lambda_0^3}+
\frac{M}{2\pi^2}\widehat C_{1}\widehat C_{2}\left(
k'^2+k^2+A\;p^2+\frac{4}{3}\frac{M\Lambda^3}{2\pi^2}V_0\right)
p^2\frac{V_0(p,\Lambda)^2}{\Lambda_0^3},
\end{equation}
where $A$ is a constant of integration which is not fixed
by the boundary conditions.  
This unfixed term arises from the solution of the homogeneous part of the
linearised RG equations, and has the same structure as the contribution
proportional to $\widehat{C_3}$.  In the above expression we have arbitrarily 
chosen to associate it with the momentum dependent perturbation multiplying  
$\widehat C_{1}\widehat C_{2}$. 

Both of these perturbations contribute to the shape parameter $P$ which is the 
observable obtained from the coefficient of $p^4$ in the effective range 
expansion.  To avoid spoiling the one to one correspondence
between observables and terms in the potential, it is possible to 
choose $A=-2$ which ensures that the contribution of 
$\widehat C_{1}\widehat C_{2}$ to $P$ vanishes.

So long as analogous procedures can be carried out to all orders, the 
effective theory defined by an expansion around this fixed point is 
systematic and completely equivalent to the effective range expansion.
This corresponds to the fact that the parts of our potential which now
contribute to observables act like a quasipotential: an energy-dependent 
boundary condition on the logarithmic derivative of the wave function at 
the origin.  Such an equivalence has previously been suggested by 
van Kolck \cite{uvk}. 

As has already been pointed out, the form of the fixed point potential 
(\ref{eq:potfp.ere}) depends on the type of cut-off used, but for small 
$p$ it always behaves like $1/\Lambda$.  We can therefore compare 
the terms that appear in the expansion of our potential (\ref{eq:pot2}) 
with those of KSW \cite{ksw2}, since the subtraction scale $\mu$
in PDS plays a similar role to $\Lambda$ in a cut-off scheme.
For example, in the case $1/a=0$ (on the critical surface), 
the lowest order contribution to their potential, $C_0(\mu)$, scales like 
$1/\mu$.
This is in agreement with the scaling behaviour of our fixed point potential.
Similarly, the term which gives the leading contribution to their
effective range, $C_2(\mu)$, scales like $1/\mu^2$ which is consistent with the
$V_0^2$ multipling $p^2$ in (\ref{eq:pot2}) and so on.  In fact, the power 
counting for perturbations around the nontrivial fixed point agrees with that 
of KSW if, as before, we assign them an order $d=\nu-1$. 

One detail remains:  the fixed point is unstable and, since $1/a\neq 0$, as 
$\Lambda\rightarrow0$ the potential will either flow to the non-trivial
fixed point or diverge to infinity.  So long as it is possible to
choose the cut-off so that $1/a<\!<\Lambda<\!<\Lambda_0$ however, the
behaviour of the potential is dominated by the flow towards the nontrivial
fixed point, and the eigenfunctions found above still define a systematic 
expansion of the potential, as noted in Ref.\cite{ksw2}.
\end{section}
\begin{section} {SUMMARY} 
In the case of weak scattering, the effective
field theory originally suggested by Weinberg \cite{wein2} to
describe NN scattering is systematic, 
and gives predictions which are independent of the regularisation scheme
used (coordinate space or momentum space cut-off, dimensional
regularisation ($\overline{MS}$), etc.).  This can be understood
in terms of an expansion of the potential around the trivial fixed point 
of the renormalisation group equation.

In the presence of a resonance or bound state close to threshold, 
conventional regularisation schemes do not lead 
to power counting schemes which are both systematic and useful.  Recently,
several equivalent power counting schemes based on new 
regularisation or subtraction schemes have been proposed which appear to avoid 
the problems mentioned above.  These correspond to an expansion of the
potential around the simplest non trivial fixed point of the
renormalisation group.  Terms in this potential have a one to one 
correspondence with on-shell scattering observables in the effective
range expansion to all orders considered so far.
The success of the effective range
expansion\cite{ere,bl} can be therefore be understood in terms of an 
effective field theory based on this nontrivial fixed 
point\cite{bvk,pkmr,ksw2}.

\end{section}

\acknowledgments
We are grateful to D. Phillips and U. van Kolck for helpful discussions. MCB
thanks the organisers of the Caltech/INT workshop which provoked this line of
investigation.  KGR would like to thank the organisers of NuSS'98 for the
opportunity to attend the school and give this talk. This work was supported 
by the EPSRC and PPARC.

\end{document}